\documentclass[fleqn]{article}
\usepackage[cp1251]{inputenc}
\usepackage[russian,english]{babel}
\usepackage{amsmath,amssymb}
\usepackage{xcolor}

\newtheorem{stat}{Statement}

\newcommand{\MI}{$\mathfrak{M}_1$}
\newcommand{\MII}{$\mathfrak{M}_2$}
\newcommand{\MO}{$\mathfrak{M}_0$}
\newcommand{\MOO}{$\mathfrak{M}_{00}$}
\newcommand{\Eef}{E_{e\!f\!f}}
\newcommand{\Fig}[3]{%
\begin{center}
\parbox{8cm}{%
\refstepcounter{figure}\includegraphics[width=8cm,height=#2cm]{#1} \noindent Figure \thefigure:\quad
#3}\end{center}}

\usepackage{amsfonts,amssymb,cite}
\usepackage{graphicx}

\def\noi{\noindent}

\newcommand{\Title}[1]{\noi {{\Large\bf #1}}\\[1ex]}

\newcommand{\Author}[2]{\noi{\bf #1}\\[2ex]\noi{\normalsize\it #2}\\}

\newcommand{\Abstract}[1]{\vskip 2mm \begin{center}
        \parbox{16.4cm}{\small\noi #1} \end{center}\medskip}
\newcommand{\foom}[1]{\protect\footnotemark[#1]}

\def\nqq{\hspace*{-2em}}

\def\Jl#1#2{#1 {\bf #2},\ }

\def\ApJ#1 {\Jl{Astroph. J.}{#1}}
\def\CQG#1 {\Jl{Class. Quantum Grav.}{#1}}
\def\DAN#1 {\Jl{Dokl. AN SSSR}{#1}}
\def\GC#1 {\Jl{Grav. Cosmol.}{#1}}
\def\GRG#1 {\Jl{Gen. Rel. Grav.}{#1}}
\def\JETF#1 {\Jl{Zh. Eksp. Teor. Fiz.}{#1}}
\def\JETP#1 {\Jl{Sov. Phys. JETP}{#1}}
\def\JHEP#1 {\Jl{JHEP}{#1}}
\def\JMP#1 {\Jl{J. Math. Phys.}{#1}}
\def\NPB#1 {\Jl{Nucl. Phys. B}{#1}}
\def\NP#1 {\Jl{Nucl. Phys.}{#1}}
\def\PLA#1 {\Jl{Phys. Lett. A}{#1}}
\def\PLB#1 {\Jl{Phys. Lett. B}{#1}}
\def\PRD#1 {\Jl{Phys. Rev. D}{#1}}
\def\PRL#1 {\Jl{Phys. Rev. Lett.}{#1}}

\def\lal{&&\nqq {}}

\def\beq{\begin{equation}}
\def\eeq{\end{equation}}
\def\bear{\begin{eqnarray}}
\def\bearr{\begin{eqnarray} \lal}
\def\ear{\end{eqnarray}}
\def\earn{\nonumber \end{eqnarray}}

\def\e{{\,\rm e}}

\begin{document}
\thispagestyle{empty}
\twocolumn[

\vspace{1cm}

\Title{Cosmological evolution of a statistical system of degenerate scalar charged fermions with an asymmetric scalar doublet. II. One-component system of doubly charged fermions\foom 1}

\Author{Yu. G. Ignat'ev$^1$, A.A. Agathonov$^2$ and D. Yu. Ignatyev$^1$}
    {$^1$Institute of Physics, Kazan Federal University, Kremlyovskaya str., 18, Kazan, 420008, Russia\\
    $^2$Lobachevsky Institute of Mathematics and Mechanics, Kazan Federal University, Kremlyovskaya str., 18, Kazan, 420008, Russia}


\Abstract
 {Based on the previously formulated mathematical model of a statistical system with scalar interaction of fermions, a cosmological model based on a one-component statistical system of doubly scalar charged degenerate fermions interacting with an asymmetric scalar doublet-canonical and phantom scalar fields-is studied. The connection of the presented model with previously studied models based on one-component and two-component fermion systems is investigated. The asymptotic and limiting properties of the cosmological model are investigated, it is shown that among all models there is a class of models with a finite lifetime. The asymptotic behavior of models near the corresponding singularities is investigated, a qualitative analysis of the corresponding dynamical system is carried out, and numerical implementations of such models are constructed. Based on numerical integration, it is shown that in the presented model there can be transitions from a stable asymptotically vacuum state with a zero canonical field and a constant phantom field corresponding to the phase of cosmological compression to a symmetric state corresponding to the expansion phase. The time interval of the transition between phases is accompanied by oscillations of the canonical scalar field.
}
\bigskip

] 
\section*{Introduction}
In recent years, due to the simultaneous direct experimental detection in 2016, gravitational waves and Black Holes \cite{GW-open1}, \cite{GW-open2} and their intensive study, which, in particular, confirmed earlier indirect observations of the orbits of stars of a supermassive Black Hole in the center of our Galaxy with a mass of about $4\cdot10^6M_\odot$(see, for example, \cite{SMBH1}, \cite{SMBH2}), as well as the existence of supermassive black holes in the centers of galaxies\footnote{Basically, quasars, such as, for example, the supermassive Black Hole SDS J140821. 67+025733.2 in the center of the quasar SDSS J140821, having a mass of $1.96\cdot 10^{11}M_\odot$.} with masses in the range $10^9\div11^{10}M_\odot$.

It is believed that supermassive Black Holes with a mass of $\sim10^9M_\odot$ Suns are the central objects of luminous quasars observed at $z>6$, but their astrophysical origin remains not fully understood. Currently, more than 200 quasars with $z>6$ and several objects with $z > 7$have been discovered. The quasar with the largest redshift at $z=7.5$, which corresponds to the age of the Universe at 650 million years, has an absolute luminosity of $1.4\cdot10^{47}$ erg/sec, while the mass estimate from the gas velocity in the quasar gives the value of $1.6\pm0,4\cdot10^9 M_\odot$ \cite{Fan}. Other detected quasars at $z> 7$ have supermassive Black Holes of similar mass. These observational data raise the question of the mechanism of formation and rapid growth of such objects in the early universe.

The results of numerical simulation impose a number of restrictions on the parameters of the formation of supermassive black holes. It was shown by \cite{Trakhtenbrot} that light embryos with a mass of $M\leqslant 10^3 M_\odot$ cannot grow to masses of the order of $10^8 M_\odot$ by $z = 6$ even with supercritical accretion. The formation of the first supermassive Black Holes with masses of $10^8 \div 10^9 M_\odot$ requires heavier nuclei of $M\sim 10^4\div 10^6 M_\odot$ and gas-rich galaxies containing quasars. However, there are currently no convincing models for the appearance of such heavy embryos. In addition, it was found that the spatial density of luminous quasars decreases rapidly with increasing redshift, and this trend increases beyond $z = 5\div 6$ \cite{Zhu}, so that such quasars are the rarest objects detected at a large redshift.

The interest in the mechanisms of formation of supermassive Black Holes, taking into account the fact of their dominant presence in the composition of quasars, is caused, in particular, by the fact that such Black Holes are formed in the composition of quasars at fairly early stages of the evolution of the Universe, before the formation of stars. This circumstance, in particular, opens up the possibility of the formation of supermassive Black Holes in conditions where scalar fields and baryonic dark matter can have a significant influence on this process. In this regard, we note the works \cite{Supermass_BH}, \cite{Shadows} and \cite{Soliton}, which consider the possibility of the existence of scalar gallos and scalar hairs in the vicinity of supermassive Black Holes.

In the works \cite{Ignat_Agaf_16_3} and \cite{Ignat_Sasha_G&G}, the following assumption was formulated based on the study of systems of scalar charged degenerate fermions with phantom interaction: 1). a cold completely degenerate Fermi system with very large effective masses of scalar charged fermions can become a good model of dark matter; 2). at a certain stage of cosmological evolution, gravitational instabilities in non-relativistic matter can lead to the appearance of isolated regions with dark matter; 3). standard Cooper mechanisms in Fermi systems with particle attraction can lead to the formation of bosons from pairs of fermions and, thereby, to the superfluidity of dark matter regions; 4). with the growth of the effective masses of fermions in the growing scalar field above the Planck value, massive fermions can form stable primary Black holes, taking into account the Hawking theorems about Black Holes, in the variant with superfluid quasi-bosons with zero spin.

In \cite{GC_21_2}, the assumption about the instability of short-wave perturbations was confirmed, but for Fermi systems with canonical scalar interaction. These preliminary studies have shown the need for a comprehensive and more in-depth study of statistical systems of scalar charged particles.

Further, in \cite{TMF_print21}, two simplest models of the interaction of fermions with an asymmetric scalar doublet were proposed: in the first model, such interaction is carried out by two types of different-grade fermions, one of which is the source of a canonical scalar field, and the second is a phantom field (model \MI); in the second model, there is one kind of fermions with a pair charge -- canonical and phantom (model \MII). A qualitative analysis of the dynamic systems of the \MI model was also carried out there. In \cite{Ignat_GC21}, a numerical simulation of the \MI cosmological model was carried out, on the basis of which the features of this model were revealed, in particular, the possibility of the existence of phases of cosmological compression, oscillations of the Hubble parameter, as well as the possibility of universes with a finite lifetime.

In this paper, we investigate the \MII\ model of a one-component statistical system with an asymmetric scalar interaction of fermions and compare it with the results of previous studies. Due to the large number of parameters of the \MII model, as well as based on aesthetic considerations, we will limit ourselves to the case of zero seed mass of $\zeta$-fermions and identify the most typical cases of the behavior of the cosmological model.

\section{Mathematical model of a cosmological system of scalar charged fermions with an asymmetric scalar Higgs doublet}
Consider the spatially flat model of the Friedman universe\footnote{In this article, we use a metric with the signature $(---+)$, the Ricci tensor is determined by convolution of the first and third indices of the curvature tensor (see, for example, \cite{Land_Field})}:
\begin{equation}\label{ds}
ds^2=dt^2-a^2(t)(dx^2+dy^2+dz^2).
\end{equation}
Let's introduce the Hubble parameter next $H(t)$
\begin{equation}\label{H}
H=\frac{\dot{a}}{a}
\end{equation}
and invariant cosmological acceleration $\Omega(t)$
\begin{equation}\label{Omega}
\Omega=\frac{\ddot{a}a}{\dot{a}^2}\equiv 1+\frac{\dot{H}}{H^2}\equiv -\frac{1}{2}(1+3w),
\end{equation}
where $w$ is the barotrope coefficient
\begin{equation}\label{w}
w=p/\varepsilon;
\end{equation}
$p$ and $\varepsilon$ -- the total pressure and energy density of cosmological matter.

Consider a statistical system consisting of $n$ varieties of degenerate scalar-charged fermions with scalar charges $q_{(a)}^r$ with respect to $N$ scalar fields $\Phi_r$. Let then
\begin{equation}\label{m_*(pm)}
m_{(a)}=m^0_{(a)}+\sum\limits_r q_{(a)}^r\Phi_r
\end{equation}
be the dynamic masses of these fermions \cite{TMF_print21} and $L_s$ is the Lagrange function of non-interacting scalar Higgs fields $L_s=\sum\limits_r L_{(r)}$
\begin{eqnarray} \label{Ls}
L_s=\frac{1}{16\pi}\sum\limits_r(e_r g^{ik} \Phi_{r,i} \Phi _{r,k} -2V_r(\Phi_r)),
\end{eqnarray}
\begin{eqnarray}
\label{Higgs}
V_r(\Phi_r)=-\frac{\alpha_r}{4} \left(\Phi_r^{2} -\frac{m_r^{2} }{\alpha_m}\right)^{2}
\end{eqnarray}
-- the potential energy of the corresponding scalar fields, $\alpha_r$ -- constants of their self-action, $m_r$ -- their quantum masses, $e_r=\pm 1$ -- indicators (the ``+'' sign corresponds to canonical scalar fields, the ``-'' sign corresponds to phantom ones).

Further, the energy-momentum tensor of scalar fields with respect to the Lagrange function \eqref{Ls} is:
\begin{eqnarray}\label{T_s}
\!\!T^i_{(s)k}=\!\frac{1}{16\pi }\sum\limits_r\bigl(2\e_r\Phi^{,i}_{r} \Phi _{r,k} -e_r\delta^i_k\Phi _{r,j} \Phi _r^{,j}
 \nonumber\\+2V_r(\Phi_r)\delta^i_k \bigr),
\end{eqnarray}
and the energy-momentum tensor of an equilibrium statistical system is equal to:
\begin{equation}\label{T_p}
T^i_{(p)k}=(\varepsilon_p+p_p)u^i u_k-\delta^i_k p,
\end{equation}
where $u^i$ is the vector of the macroscopic velocity of the statistical system, $\varepsilon_p$ and $p_p$ are its energy density and pressure. Einstein's equations for the system ``scalar fields+ particles'' have the form:
\begin{equation}\label{Eq_Einst_G}
R^i_k-\frac{1}{2}\delta^i_k R=8\pi T^i_k+ \delta^i_k \Lambda_0,
\end{equation}
where
\[T^i_k=T^i_{(s)k}+^i_{(p)k},\]
$\Lambda_0$ -- the seed value of the cosmological constant associated with its observed value $\Lambda$ by the relation:
\begin{equation}\label{lambda0->Lambda}
\Lambda=\Lambda_0-\frac{1}{4}\sum\limits_r \frac{m^4_r}{\alpha_r}.
\end{equation}
\subsection{General relations for models of degenerate fermions with scalar interaction}
One can prove\footnote{For a single-component system of degenerate fermions, see \cite{Ignat_Agaf_Dima14_3_GC}, \cite{Ignat15_2}, and for a multicomponent system, see \cite{TMF_print21}}, that a strict consequence of the general relativistic kinetic theory for statistical systems of completely degenerate fermions is the Fermi momentum conservation law  $\pi_{(a)}$ of each component
\begin{equation}\label{ap}
a(t)\pi_{(a)}(t)=\mathrm{Const}.
\end{equation}
Assuming further for certainty $a(0)=1$ (see \cite{Ignat_GC21}) and
\begin{equation}\label{a-xi}
\xi=\ln a;\quad \xi\in(-\infty,+\infty); \quad \xi(0)=0,
\end{equation}
we introduce dimensionless functions
\begin{equation}\label{psi0}
\psi_{(a)}=\frac{\pi^0_{(a)} \mathrm{e}^{-\xi}}{|m_{(a)}|}, \quad (\pi^0_{(a)}=\pi_{(a)}(0)),
\end{equation}
equal to the ratio of the Fermi momentum $\pi_{(a)}$ to the total energy of the fermion, as well as the functions $F_1(\psi)$ and $F_2(\psi)$:
\begin{equation}\label{F_1}
F_1(\psi)=\psi\sqrt{1+\psi^2}-\ln(\psi+\sqrt{1+\psi^2});
\end{equation}
\begin{equation}
\label{F_2}
F_2(\psi)=\psi\sqrt{1+\psi^2}(1+2\psi^2)-\ln(\psi+\sqrt{1+\psi^2}),
\end{equation}
with the help of which we will determine the macroscopic scalars of the statistical system:
density of the number of particles of the variety $a$ --
\begin{equation}\label{2_3}
n^{(a)}=\frac{1}{\pi^2}\pi_{(a)}^3;
\end{equation}
pressure of the $a$-th component --
\begin{equation}\label{2_3a}
{\displaystyle
\begin{array}{l}
\varepsilon_p = {\displaystyle\sum\limits_a\frac{m_{(a)}^4}{8\pi^2}}F_2(\psi_{(a)});
\end{array}}
\end{equation}
the energy density of the fermion system --
\begin{equation}\label{2_3b}{\displaystyle
\begin{array}{l}
p_p ={\displaystyle\sum\limits_a\frac{m_{(a)}^4}{24\pi^2}}(F_2(\psi_{(a)})-4F_1(\psi_{(a)}))
\end{array}}
\end{equation}
and the density of the scalar charge of the fermion system with respect to the scalar field $\Phi_r$ --
\begin{equation}\label{2_3c}{\displaystyle
\begin{array}{l}
\sigma_r={\displaystyle \sum\limits_a q^r_{(a)}\frac{m_{(a)}^3}{2\pi^2}}F_1(\psi_{(a)}).
\end{array}}
\end{equation}
We will also write out useful expressions for further derived functions $F_1(x)$ and $F_2(x)$:
\begin{eqnarray}\label{F'_12}
F'_1(x)=\frac{2x^2}{\sqrt{1+x^2}};\quad F'_2(x)=8x^2\sqrt{1+x^2}.
\end{eqnarray}
We also note a useful relation
\begin{equation}\label{E_P_f}
\varepsilon_p+p_p\equiv \frac{1}{3\pi^2}\sum\limits_a m^4_{(a)}\psi_{(a)}^3\sqrt{1+\psi_{(a)}^2}.
\end{equation}

Thus, in the cosmological model consisting of a system of degenerate scalar charged fermions and scalar fields, all macroscopic scalars are determined by explicit algebraic functions of scalar potentials $\Phi_1(t),\ldots,\Phi_N(t)$.

A complete closed normal autonomous system of ordinary differential equations describing a cosmological model based on a statistical system of completely degenerate scalar charged fermions has the form (see \cite{Ignat20_1}, \cite{TMF_print21}):
\begin{eqnarray}
\label{dot(xi)}
\dot{\xi}=H;\\
\label{dot(H)}
\dot{H}=-\sum\limits_r\frac{ e_r Z^2_r}{2}-\sum\limits_a \frac{4m_{(a)}^2\psi_{(a)}^3}{3\pi}\sqrt{1+\psi_{(a)}^2};\\
\label{dot(Phi)}
\dot{\Phi_r}=Z_r, \qquad (r=\overline{1,N});\\
\label{dot(Z)}
e_r\dot{Z}_r=-e_r 3HZ_r- m^2_r\Phi_r+\alpha_r\Phi^3_r-8\pi\sigma_r(t).
\end{eqnarray}

A strict consequence of the dynamical system \eqref{dot(xi)} -- \eqref{dot(Z)} is the total energy integral $3H^2-8\pi\Eef=\mathrm{Const}$, the partial zero value of which corresponds to the Einstein equation $^4_4$
\begin{eqnarray}\label{Surf_EinstNn}
3H^2-\Lambda-\sum\limits_r \left(\frac{e_rZ^2}{2}-\frac{m^2_r\Phi^2_r}{2}+\frac{\alpha_r\Phi^4_r}{4}\right)\nonumber\\
-\frac{1}{\pi}\sum\limits_a m^4_{(a)}F_2(\psi_{(a)} \equiv 3H^2-8\pi \Eef =0.
\end{eqnarray}
The equations \eqref{Surf_EinstNn} defines a certain hypersurface $\Sigma_E$ in the $2N+2$-dimensional phase space of the dynamical system \eqref{dot(xi)} -- \eqref{dot(Z)}. In the future, we will call this hypersurface the Einstein hypersurface. The points of the phase space $\mathbb{R}_{2N+2}$ in which the effective energy $\Eef$  is negative are inaccessible to the dynamical system. The inaccessible region is separated from the accessible region of the phase space by a hypersurface of zero effective energy $S_{E}\subset \mathbb{R}_{2N+2}$, which is a cylinder with the axis $OH$:
\begin{eqnarray}\label{S_ENn}
8\pi\Eef\equiv \Lambda+\frac{1}{\pi}\sum\limits_a m^4_{(a)}F_2(\psi_{(a)}\nonumber\\
+\sum\limits_r \left(\frac{e_rZ^2}{2}+\frac{m^2_r\Phi^2_r}{2}-\frac{\alpha_r\Phi^4_r}{4}\right)
=0,
\end{eqnarray}
moreover, the hypersurface of zero effective energy \eqref{S_ENn} concerns the Einstein hypersurface \eqref{Surf_EinstNn} in the hyperplane
$H=0$:
\begin{equation}\label{H=0}
\Sigma_E \cap S_E =H=0.
\end{equation}
The fact that the Einstein equation \eqref{Surf_EinstNn} is the quotient of the first integral of the dynamical system \eqref{dot(xi)} -- \eqref{dot(Z)}, allows one to use it to determine the initial value of the Hubble parameter $H_0=H(0)$, which we will do in future. The equation \eqref{Surf_EinstNn} is square with respect to $H(t)$, so it has two symmetric roots $\pm H(t)$, and the positive root corresponds to the expansion of the universe, and the negative root corresponds to compression.

\subsection{Cosmological model \MII\ with scalar interaction of a one-component fermion system}
Further, $L_s$ is the Lagrange function of interacting of canonical ($\Phi$) and phantom ($\varphi$) scalar fields
\begin{eqnarray} \label{Ls}
L_s= \frac{1}{16\pi}(g^{ik} \Phi_{,i} \Phi _{,k} -2V(\Phi))\nonumber\\
+\frac{1}{16\pi}(- g^{ik} \varphi_{,i} \varphi _{,k} -2\mathcal{V}(\varphi)),
\end{eqnarray}
where
\begin{eqnarray}
\label{Higgs}
\!\!\!\!V(\Phi)\!=\!-\frac{\alpha}{4} \left(\Phi^{2}\! -\frac{m^{2} }{\alpha}\right)^{2}\!;
\mathcal{V}(\varphi)\!=\!-\frac{\beta}{4} \left(\varphi^2\! -\frac{\mathfrak{m}^2}{\beta}\right)^{2}\nonumber\
\end{eqnarray}
-- the potential energy of the corresponding scalar fields, $\alpha$ and $\beta$ -- the constants of their interaction, $m$ and $\mathfrak{m}$ -- masses of their quanta. As a carrier of scalar charges, we consider a one-component degenerate fermion system in which the fermions $\zeta$ simultaneously have two charges: the canonical charge $e$ with respect to the canonical scalar field $\Phi$ and the phantom charge $\epsilon$ with respect to the phantom field $\varphi$. In this case, fermions can have some seed mass $m_0$\footnote{see a discussion on this issue in \cite{Ignat_GC21}} and the initial Fermi impulse $\pi_0$. In this case, the formula \eqref{psi0} and \eqref{2_3c} take the form:
\begin{equation}\label{psi2}
\psi=\frac{\pi_0}{|e\Phi+\epsilon\varphi|}\mathrm{e}^{-\xi};
\end{equation}
\begin{eqnarray}\label{sigma-sigma}
\sigma_c=\frac{e}{2\pi^2}(e\Phi+\epsilon\varphi)^3 F_1(\psi);\nonumber\\
\sigma_f=\frac{\epsilon}{2\pi^2}(e\Phi+\epsilon\varphi)^3 F_1(\psi).
\end{eqnarray}
Thus, in the \MII model, the following relation takes place:
\begin{equation}\label{signa=sigma}
\epsilon\sigma_c=e\sigma_f.
\end{equation}

Let us write out a complete normal system of Einstein's equations and scalar fields $\Phi(t)$ and $\varphi(t)$ for this one-component system of scalar charged degenerate fermions. In an obviously non-singular form, the normal system of ordinary differential equations of the model under study \eqref{dot(xi)}, \eqref{dot(H)}, \eqref{dot(Phi)} -- \eqref{dot(Z)} taking into account \eqref{sigma-sigma} takes the form:
\begin{equation}\label{dxi/dt2}
\dot{\xi}=H;
\end{equation}
\begin{eqnarray}\label{dH/dt2}
\dot{H}=- \frac{Z^2}{2}+ \frac{z^2}{2}-\frac{4}{3\pi}\pi_0^3\mathrm{e}^{-3\xi}\times\nonumber\\
\sqrt{(e\Phi+\epsilon\varphi)^2+\pi^2_0\mathrm{e}^{-2\xi}};
\end{eqnarray}
\begin{equation}
\label{dPhi/dt2}
\dot{\Phi}=Z; \qquad \dot{\varphi}=z;
\end{equation}
\begin{eqnarray}
\label{dZ/dt2}
\dot{Z}=\!\!-3HZ-\!\! m^2\Phi +\alpha\Phi^3-\frac{4e\pi_0}{\pi}(e\Phi+\epsilon\varphi)\mathrm{e}^{-\xi} &\nonumber\\
\times \sqrt{\pi^2_0 \mathrm{e}^{-2\xi}+(e\Phi+\epsilon\varphi)^2}+\frac{4e}{\pi}(e\Phi+\epsilon\varphi)^3 & \nonumber\\
\times\ln\biggl(\frac{\pi_0\mathrm{e}^{-\xi}+\sqrt{\pi^2_0 \mathrm{e}^{-2\xi}+(e\Phi+\epsilon\varphi)^2}}{\sqrt{(e\Phi+\epsilon\varphi)^2}}\biggr);&
\end{eqnarray}
\begin{eqnarray}
\label{dz/dt2}
\dot{z}=-3Hz+\mathfrak{m}^2\varphi -\beta\varphi^3+\frac{4\epsilon\pi_0}{\pi}(e\Phi+\epsilon\varphi)\mathrm{e}^{-\xi}\nonumber\\
\times\sqrt{\pi^2_0 \mathrm{e}^{-2\xi}+(e\Phi+\epsilon\varphi)^2}-\frac{4\epsilon}{\pi}(e\Phi+\epsilon\varphi)^3\nonumber\\
\times\ln\biggl(\frac{\pi_0\mathrm{e}^{-\xi}+\sqrt{\pi^2_0 \mathrm{e}^{-2\xi}+(e\Phi+\epsilon\varphi)^2}}{\sqrt{(e\Phi+\epsilon\varphi)^2}}\biggr).
\end{eqnarray}
The first integral of the system of equations \eqref{Surf_EinstNn} for the \MII model takes the form:

\begin{eqnarray}\label{Surf_Einst2}
\!\!\!\!\! 3H^2-\Lambda-\frac{Z^2}{2}+\frac{z^2}{2}+\frac{m^2}{2}\Phi^2-\frac{\alpha\Phi^4}{4}+\frac{\mathfrak{m}^2}{2}\varphi^2 &
\nonumber\\
-\frac{\beta\varphi^4}{4}
-\frac{\mathrm{e}^{-\xi}}{\pi}\pi_0\sqrt{\pi^2_0 \mathrm{e}^{-2\xi}+(e\Phi+\epsilon\varphi)^2}&\nonumber\\
\times\bigl(2\pi^2_0\mathrm{e}^{-2\xi}+(e\Phi+\epsilon\varphi)^2 \bigr)+\frac{(e\Phi+\epsilon\varphi)^4}{\pi} &\nonumber\\
\times\ln\biggl(\frac{\pi_0\mathrm{e}^{-\xi}+\sqrt{\pi^2_0 \mathrm{e}^{-2\xi}+(e\Phi+\epsilon\varphi)^2}}{\sqrt{(e\Phi+\epsilon\varphi)^2}}\biggr)=0.&
\end{eqnarray}
Thus, the complete normal autonomous system of ordinary differential equations of the \MII model consists of the equations \eqref{dxi/dt2} -- \eqref{dZ/dt2} and \eqref{dz/dt2}. Moreover, we have  the total energy integral \eqref{Surf_Einst2} as the first integral of this system, which represents the Einstein hypersurface equation $\Sigma_E$ in the 6-dimensional phase space of the dynamical system $\mathbb{R}_6=\{\xi,H,\Phi,Z,\varphi,z\}$, on which all the phase trajectories of this system lie.

\subsection{Relation of mathematical models \MII, \MI\ and \MOO\ for scalar singlets\label{sub_srav}}
Following \cite{Ignat_GC21}, we will call the cosmological model with a one-component system of scalar charged fermions with a scalar singlet with a quadratic interaction potential, a non-negative dynamic mass and a non-negative Hubble parameter $H\geqslant0$, studied in \cite{Ignat15_2}, \cite{Ignat_Sasha_G&G}, as the \MOO model, a model with a one-component system of scalar charged fermions with a scalar Higgs singlet and an arbitrary sign of the Hubble parameter, investigated in \cite{Ignat20_2}, as the \MO \ model and, finally,a model with a two-component system of single scalar charged fermions, studied in the work \cite{Ignat_GC21} as the \MI model\footnote{Singlets with both canonical and phantom scalar fields were studied in the \MOO\ and \MO\ models.}.

The dynamic equations of the model \MII\ \eqref{dxi/dt2} -- \eqref{Surf_Einst2} in the case of one of the scalar singlets, canonical, \textbf{C}, or phantom, \textbf{F}, transform into the corresponding equations of the two-component model \MI, studied in the previous part of the work \cite{Ignat_GC21}, for the following parameter values and initial conditions:
\begin{eqnarray}
\mathfrak{M}_2 \rightleftarrows \mathfrak{M}_1:\hskip 3cm\nonumber\\
\label{C:MII->MI}
\!\!\!\mathbf{C}: \epsilon=0; \varphi(0)=0; z(0)=0; \pi_f=0; \pi_c=\pi_0;\\
\label{F:MII->MI}
\!\!\!\mathbf{F}: e=0; \Phi(0)=0; Z(0)=0; \pi_c=0; \pi_f=\pi_0.
\end{eqnarray}

Further, the dynamic equations of the model \MII\ \eqref{dxi/dt2} -- \eqref{Surf_Einst2} in the case of one of the scalar singlets, canonical, \textbf{C}, or phantom, \textbf{F}, transform into the corresponding equations of the one-component model \MO, studied in \cite{Ignat20_2}, for the following parameter values and initial conditions:
\begin{eqnarray}
\mathfrak{M}_2 \rightleftarrows \mathfrak{M}_0:\hskip 3cm\nonumber\\
\label{C:MII->M0}
\!\!\!\mathbf{C}: \epsilon=0; \varphi(0)=0; z(0)=0; \pi_f=0; \pi_c=\pi_0;\\
\label{F:MII->M0}
\!\!\!\mathbf{F}: e=0; \Phi(0)=0; Z(0)=0; \pi_c=0; \pi_f=\pi_0.
\end{eqnarray}

Finally, the dynamic equations of the \MII\ \eqref{dxi/dt2} -- \eqref{Surf_Einst2} model in the case of one of the scalar singlets, canonical, \textbf{C}, or phantom, \textbf{F}, are equivalent to the dynamic equations of the \MOO model studied in \cite{Ignat15_2}, only under the following parameter values and conditions:
\begin{eqnarray}
\mathfrak{M}_2 \rightleftarrows \mathfrak{M}_{00}:\hskip 3cm\nonumber\\
\label{CondH>0}
H\geqslant 0;\; \Phi\geqslant0;\;\varphi\geqslant0;\; \alpha=0;\; \beta=0;\hskip 1.3cm\\
\label{C:MII->M00}
\!\!\!\mathbf{C}: \epsilon=0; \varphi(0)=0; z(0)=0; \pi_f=0; \pi_c=\pi_0;\\
\label{F:MII->M00}
\!\!\!\mathbf{F}: e=0; \Phi(0)=0; Z(0)=0; \pi_c=0; \pi_f=\pi_0.
\end{eqnarray}
\section{Qualitative analysis and asymptotic behavior of the \MII model}
\subsection{Singular points of the dynamic system of the \MII model}
The singular points of a dynamical system are determined by the equality of the right parts of the normal system of equations to zero. Thus, we obtain from \eqref{dxi/dt2}, \eqref{dH/dt2}, \eqref{dPhi/dt2}, \eqref{dZ/dt2} and \eqref{dz/dt2} a system of algebraic equations for finding the coordinates of singular points:
\begin{equation}
\label{Zz=02}
Z=0;\quad z=0;
\end{equation}
\begin{equation}\label{H=02}
H=0;
\end{equation}
\begin{equation}
\label{-3HZ2}
-m^2\Phi +\alpha\Phi^3-\frac{4e}{\pi}(e\Phi+\epsilon\varphi)^3 F_1(\psi)=0;
\end{equation}
\begin{equation}\label{-3Hz2}
\mathfrak{m}^2\varphi -\beta\varphi^3 +\frac{4\epsilon}{\pi}(e\Phi+\epsilon\varphi)^3 F_1(\psi)=0;
\end{equation}
\begin{equation} \label{dotH=02}
-\frac{4}{3\pi}(\pi_0)^3\mathrm{e}^{-3\xi}\sqrt{(e\Phi+\epsilon\varphi)^2+\pi^2_0\mathrm{e}^{-2\xi}}=0.
\end{equation}
In addition, one must take into account the integral of the total energy \eqref{Surf_Einst2}, -- the coordinates of the singular point must satisfy this equation, which, taking into account \eqref{Zz=02}, takes the form:
\begin{eqnarray}\label{Surf_02}
-\Lambda+\frac{m^2}{2}\Phi^2-\frac{\alpha\Phi^4}{4}+\frac{\mathfrak{m}^2}{2}\varphi^2-\frac{\beta\varphi^4}{4}\nonumber\\
-\frac{1}{\pi}(e\Phi+\epsilon\varphi)^4 F_2(\psi)=0.
\end{eqnarray}

The equation \eqref{dotH=02} has a unique solution: $\xi=+\infty$. Taking into account that for $\xi\to+\infty$ $F1(\psi)=F2(\psi)=0$, we obtain the independent equations:
\begin{equation}\label{EqFf}
m^2\Phi-\alpha\Phi^3=0; \quad \mathfrak{m}^2\varphi-\beta\varphi^3=0.
\end{equation}
Thus, ($\xi=+\infty$) leads us to solutions of the \MII model
\begin{equation}\label{M_pm_pm2}
\!\!\! M_{\infty,\pm,\pm}\!\!=\!\!\biggl(\infty,0,\pm\frac{m}{\sqrt{\alpha}},0,\pm\frac{\mathfrak{m}}{\sqrt{\beta},0}\biggr),\; \Lambda_0=0;
\end{equation}
\begin{equation}\label{M_pm_02}
\!\!\! M_{\infty,\pm,0}\!\!=\!\!\biggl(\infty,0,\pm\frac{m}{\sqrt{\alpha}},0,0,0\biggr),\; \Lambda_0=\frac{\mathfrak{m}^4}{4\beta};
\end{equation}
\begin{equation}\label{M_0_pm2}
\!\!\! M_{\infty,0,\pm}\!\!=\!\!\biggl(\infty,0,0,0,\pm\frac{\mathfrak{m}}{\sqrt{\beta}},0\biggr),\; \Lambda_0=\frac{m^4}{4\alpha};
\end{equation}

\begin{stat}\label{stat_sing_points2}
The dynamical system \eqref{dxi/dt2} -- \eqref{Surf_Einst2} has singular points only at special values of the cosmological constant $\Lambda_0$:\\
\noindent$\bullet$ at $\Lambda_0=0$ $\biggl(\displaystyle\Lambda=-\frac{m^4}{4\alpha}-\frac{\mathfrak{m}^4}{4\beta}\biggr)$ -- \\
4 singular points in the infinite future $M_{\infty,\pm,\pm}$ \eqref{M_pm_pm2} with non-zero values of scalar potentials;\\
\noindent$\bullet$ at $\displaystyle\Lambda_0=\frac{\mathfrak{m}^4}{4\beta}$ $\displaystyle\biggl(\Lambda=-\frac{m^4}{4\alpha}\biggr)$ -- \\
2 singular points in the infinite future $M_{\infty,\pm,0}$ \eqref{M_pm_02} with a zero value of the phantom field potential;\\
\noindent$\bullet$ at $\displaystyle\Lambda_0=\frac{m^4}{4\alpha}$ $\displaystyle\biggl(\Lambda=-\frac{\mathfrak{m}^4}{4\beta}\biggr)$ -- \\
2  singular points in the infinite future $M_{\infty,0,\pm}$ \eqref{M_0_pm2} with a zero value of the potential of the canonical field.

In the case of an arbitrary value of the cosmological constant, the dynamical system \eqref{dxi/dt2} -- \eqref{Surf_Einst2} has no singular points at all, but at special values of the cosmological constant, the dynamical system \eqref{dxi/dt2} -- \eqref{Surf_Einst2} can have either 2 or 4 singular points in the infinite future.
\end{stat}
\subsection{Limiting and asymptotic properties of the \MII model \label{ssM1}}
\subsubsection{Asymptotic properties of the model for $a\to\infty$\label{asimp_dyn8}}
Note, first, that in the absence of fermions ($\pi_0=0$), the system of equations \eqref{dxi/dt2} -- \eqref{Surf_Einst2} continuously transforms to the system of equations for the vacuum Higgs doublet (see \cite{YuKokh_TMF}):
\begin{eqnarray}
\label{dH/dt_Vac}
\dot{H}=-\frac{Z^2}{2}+\frac{z^2}{2};\\
\label{dZ/dt_Vac}
\dot{Z}=-3HZ-m^2\Phi+\alpha\Phi^3;\\
\label{dz/dt_Vac}
\dot{z}=-3Hz+\mathfrak{m}^2\varphi-\beta\varphi^3;\\
\label{SurfEinst_Vac}
3H^2-\Lambda-\frac{Z^2}{2}+\frac{z^2}{2}-\frac{m^2\Phi^2}{2}+\frac{\alpha\Phi^4}{4} \nonumber\\
-\frac{\mathfrak{m}^2\varphi^2}{2}+\frac{\beta\varphi^4}{4}=0.
\end{eqnarray}

Secondly, we note that the beginning of the Universe (the cosmological singularity $a=0$) corresponds to $\xi\to -\infty$, and the infinite future $a\to\infty$ -- $\xi\to+\infty$ (if the model allows such a state). It can be easily seen that for $\xi\to+\infty$, the system of equations \eqref{dxi/dt2} -- \eqref{Surf_Einst2} also asymptotically tends to the system of equations for the vacuum Higgs doublet \eqref{dH/dt_Vac} -- \eqref{SurfEinst_Vac}. Therefore, if the cosmological model admits the state$\xi\to+\infty$\footnote{This state is not allowed in all cases of model parameters, see below.} then the results of qualitative and numerical analysis of the vacuum model of an asymmetric scalar doublet \cite{YuKokh_TMF} can be used to study the evolution of the cosmological model at late stages.

Third, for $e=0,\Phi=0$ or $\epsilon=0,\varphi=0$, the model transforms into the model of a single-component scalar charged Fermi system with the corresponding scalar singlet \cite{Ignat20_1}.

Fourth, for both zero charges of $\zeta$-fermions, the system of equations \eqref{dxi/dt2} -- \eqref{Surf_Einst2} continuously transforms to a system of equations for a cosmological model based on a vacuum asymmetric scalar Higgs doublet and a neutral one-component Fermi liquid (ultrarelativistic at $m_0=0$).

\subsubsection{Asymptotic properties of the model near the singularity $a\to0$}
We now investigate the behavior of the cosmological model near the cosmological singularity $\xi\to-\infty$ ($a\to 0$). It follows from the equations \\eqref{dxi/dt2}, \eqref{dH/dt2}, \eqref{dPhi/dt2}, \eqref{dZ/dt2} and \eqref{dz/dt2} that such a state is always possible for $\pi_0\not\equiv 0$. In this case, for $\xi\to-\infty$, $H\to\pm\infty$,
\begin{equation}\label{exi<<pi0}
|e\Phi+\epsilon\varphi|\ll \pi_0\mathrm{e}^{-\xi}
\end{equation}
the system of equations \eqref{dxi/dt2} -- \eqref{Surf_Einst2} reduces to the following:
\begin{eqnarray}
\label{dxi/dt-dPhi_Phi_8}
\dot{\xi}=H;\qquad \dot{\Phi}=Z;\qquad \dot{\varphi}=z;\\
\label{dH/dt_M2_8}
\dot{H}=-\frac{4\mathrm{e}^{-4\xi}}{3\pi}\pi^4_0; \\
\label{dZ/dt_M2_8}
\dot{Z}=-3HZ-\frac{4e\pi^2_0\mathrm{e}^{-2\xi}}{\pi}(e\Phi+\epsilon\varphi) \ ;\\
\label{dz/dt_M2_8}
\dot{z}=-3Hz+\frac{4\epsilon\pi^2_0\mathrm{e}^{-2\xi}}{\pi}(e\Phi+\epsilon\varphi).
\end{eqnarray}
%

By replacing the variable in the equation \eqref{dH/dt_M2_8} $d/dt=H d/d\xi$, we find its solution:
\begin{equation}\label{H_xi-8}
H=\pm\sqrt{\frac{2}{3\pi}}\pi^2_0\mathrm{e}^{-2\xi},
\end{equation}
where the sign $+$ corresponds to the exit from the singularity, $-$ -- to the entrance to it. Substituting $H$ from the equation \eqref{dxi/dt-dPhi_Phi_8} into \eqref{H_xi-8}, we obtain a differential equation with respect to $\xi(t)$:
\begin{eqnarray}
\dot{\xi}=\pm \sqrt{\frac{2}{3}}\pi^2_0\mathrm{e}^{-2\xi}\Rightarrow
d\mathrm{e}^{2\xi}=\pm \sqrt{\frac{8}{3}}\pi^2_0 dt,\nonumber
\end{eqnarray}
from where, taking into account the definition of $\xi(t)$ \eqref{a-xi} and the condition for the existence of a singularity at the point $t_0:\ \xi(t_0)=-\infty$ $\Rightarrow a(t_0)=0$, we get the asymptotics near the singularity:
\begin{eqnarray}\label{a(t_0)}
\mathrm{e}^\xi\equiv a(t)\propto \displaystyle\biggl(\frac{8}{3}\biggr)^{\frac{1}{4}}\pi_0\sqrt{|t-t_0|}.
\end{eqnarray}

Thus, near the singularity $t\to t_0$, the scale factor and the Hubble parameter have the following asymptotics:
\begin{equation}\label{a(t0),H(t0)}
\left.a(t)\right|_{t\to t_0}\propto \sqrt{|t-t_0|}; \: \left.H(t)\right|_{t\to t_0}\propto \frac{1}{t-t_0}.
\end{equation}
Finally, calculating the invariant cosmological acceleration$\Omega$ \eqref{Omega}  using \eqref{dH/dt_M2_8}, \eqref{H_xi-8} and \eqref{a(t_0)}, we obtain near the singularity:
\begin{equation}\label{Omega_8}
\left.\Omega(t)\right|_{t\to t_0}\simeq -1,
\end{equation}
which corresponds, as is known, to the ultrarelativistic equation of state $w=1/3$.

Multiplying both parts of the equation \eqref{dZ/dt_M2_8} by $e$ and both parts of the equation \eqref{dz/dt_M2_8} by $\epsilon$, adding the equations obtained in this way and introducing a new field variable
\begin{equation}\label{chi}
\chi(t)\equiv e\Phi+\epsilon\varphi,
\end{equation}
we obtain a linear homogeneous differential equation for the function $\chi(t)$
\begin{equation}\label{Eq_chi}
\ddot{\chi}+3H\dot{\chi}+\frac{4\pi^2_0 e^{-2\xi}}{\pi}(e^2-\epsilon^2)\chi=0.
\end{equation}
This equation is integrated in the Bessel functions of the I-st kind $J_2(z)$ and $Y_2(z)$:\footnote{see, for example, \cite{Lebed}}
\begin{equation}\label{sol_chi}
\chi=C_1 \frac{J_2(\sqrt{\nu|t-t_0|})}{t-t_0}+C_2 \frac{Y_2(\sqrt{\nu|t-t_0|})}{t-t_0},
\end{equation}
where
\begin{equation}\label{nu}
\nu^2\equiv \frac{2\sqrt{6}\pi^2_0}{\pi}|e^2-\epsilon^2|.
\end{equation}

Substituting the found solution in the right parts of \eqref{dZ/dt_M2_8} and \eqref{dz/dt_M2_8}, we obtain asymptotic solutions for the derivatives of potentials and then, the potentials themselves. So, for example,
\[Z\simeq -\frac{\sqrt{6} e}{\pi\tau^2}\int \tau\bigl(C_1 J_2(\sqrt{\nu\tau})+C_2 Y_2(\sqrt{\nu\tau})\bigr)d\tau,\]
where $\tau=t-t_0$.
By replacing the variable $x=\sqrt{\nu\tau}$ in this integral, we bring it to the form:
\[Z\simeq -\frac{2\sqrt{6} e}{\pi\tau^2\nu^2}\int x^3\bigl(C_1 J_2(x)+C_2 Y_2(x)\bigr)dx.\]
Applying the well-known recurrence relation for Bessel functions $Z_p(x)$ (see, for example, \cite{Lebed})
\[\frac{d}{dx}x^pZ_p(x)=x^pZ_{p-1}(x),\]
we find the asymptotics for $Z(x)$:
\begin{equation}\label{Z(x)8}
Z\simeq -\frac{2\sqrt{6}e}{\pi \sqrt{\nu\tau}}\bigl(C_1 J_3(\sqrt{\nu\tau})+C_2 Y_3(\sqrt{\nu\tau})\bigr).
\end{equation}
\subsection{Eigenvalues of the dynamic system matrix}
Let us proceed to the study of the nature of the singular points of a dynamical system in those special cases I--III (p. \pageref{stat_sing_points2}) when these points exist. Calculating the main matrix of the dynamical system \eqref{dxi/dt2} -- \eqref{dz/dt2} at the singular points \eqref{Zz=02} and \eqref{H=02}, we find:
\begin{equation}\label{A(M2)}
A(M)=\displaystyle\left(\begin{array}{cccccc}
0 & 1 & 0 & 0 &0 & 0 \\
\frac{\partial P_1}{\partial \xi} & 0 & \frac{\partial P_1}{\partial \Phi} & 0 &  \frac{\partial P_1}{\partial \varphi} & 0\\
0 & 0 & 0 & 1 & 0 & 0 \\
\frac{\partial P_2}{\partial \xi} & 0 & \frac{\partial P_2}{\partial \Phi} & 0 & \frac{\partial P_2}{\partial \varphi} & 0\\
0 & 0 & 0 & 0 & 0 & 1 \\
\frac{\partial P_3}{\partial \xi} & 0 & \frac{\partial P_3}{\partial \Phi} & 0 & \frac{\partial P_3}{\partial \varphi} & 0 \\
\end{array}\right)_M,
\end{equation}
where $P_1$, $P_2$ and $P_3$ are the right-hand sides of the equations \eqref{dH/dt2},  \eqref{dZ/dt2} and \eqref{dz/dt2}, respectively. One can make sure that in the general case, the matrix \eqref{A(M2)} is not degenerate.

Using the obvious relations
\[\frac{\partial \psi}{\partial \xi}=-\psi;\; \frac{\partial \psi}{\partial \Phi_r}=-\frac{e^r\psi}{e\Phi+\epsilon\varphi}, \]
and also the definition of the functions $P_1$, $P_2$, $P_3$ and the expression \eqref{F'_12} for the derivatives of the functions $F_1(x)$ and $F_2(x)$, we obtain expressions for the partial derivatives of the functions $P_k$ included in the matrix $A(M)$:
\[\frac{\partial P_1}{\partial \xi}=\frac{4\pi_0^3\mathrm{e}^{-3\xi} }{3\pi}\frac{3(e\Phi+\epsilon\varphi)^2 + 4\pi_0^2 \mathrm{e}^{-2\xi}}{\sqrt{(e\Phi+\epsilon\varphi)^2 + \pi_0^2 \mathrm{e}^{-2\xi}}};\]
\[\frac{\partial P_1}{\partial \Phi}=-\frac{4e\pi^3_0  \mathrm{e}^{-3\xi}}{3\pi}\ \frac{e\Phi+\epsilon\varphi}{\sqrt{(e\Phi+\epsilon\varphi)^2 + \pi_0^2 \mathrm{e}^{-2\xi}}};\]
\[\frac{\partial P_1}{\partial \varphi}=-\frac{4\epsilon\pi^3_0 \mathrm{e}^{-3\xi} }{3\pi}\ \frac{e\Phi+\epsilon\varphi}{\sqrt{(e\Phi+\epsilon\varphi)^2 + \pi_0^2 \mathrm{e}^{-2\xi}}};  \]
\[\frac{\partial P_2}{\partial \xi}=\frac{8\psi^3}{\pi\sqrt{1+\psi^2}}(e\Phi+\epsilon\varphi)^3;\]
\begin{eqnarray}\frac{\partial P_2}{\partial \Phi}=-m^2+3\alpha\Phi^2-\frac{12e^2}{\pi}(e\Phi+\nonumber\\
\epsilon\varphi)^2 F_1(\psi)+\frac{8e^2 \psi\pi^2_0\mathrm{e}^{-2\xi}}{\pi\sqrt{1+\psi^2}}; \nonumber
\end{eqnarray}
\[\frac{\partial P_2}{\partial \varphi}=-\frac{12e\epsilon}{\pi}(e\Phi+\epsilon\varphi)^2 F_1(\psi)+
\frac{8e\epsilon \psi\pi^2_0\mathrm{e}^{-2\xi}}{\pi\sqrt{1+\psi^2}};\]
\[
\frac{\partial P_3}{\partial \xi}=-\frac{8\epsilon\psi^3}{\pi\sqrt{1+\psi^2}}(e\Phi+\epsilon\varphi)^3; \]
\[\frac{\partial P_3}{\partial \Phi}=\frac{12e\epsilon}{\pi}(e\Phi+\epsilon\varphi)^2 F_1(\psi)-
\frac{8e\epsilon \psi\pi^2_0\mathrm{e}^{-2\xi}}{\pi\sqrt{1+\psi^2}};\]
\begin{eqnarray}
\frac{\partial P_3}{\partial \varphi}=\mathfrak{m}^2-3\beta\varphi^2+\frac{12\epsilon^2}{\pi}(e\Phi+\epsilon\varphi)^2 F_1(\psi)\nonumber\\
-\frac{8\epsilon^2 \psi\pi^2_0\mathrm{e}^{-2\xi}}{\pi\sqrt{1+\psi^2}}. \nonumber
\end{eqnarray}

Calculating the value of these coefficients of the matrix $A(M)$ at the points $M_{+\infty}$, we obtain for non-zero coefficients of the matrix the values:
\begin{eqnarray}\label{koef}
\left.\frac{\partial P_2}{\partial \Phi}\right|_{\xi\to+\infty}=-m^2+3\alpha\Phi^2;\nonumber\\
\left.\frac{\partial P_3}{\partial \varphi}\right|_{\xi\to+\infty}=\mathfrak{m}^2-3\beta\varphi^2.\\
\end{eqnarray}
Thus, at the points $M_{\infty,\pm,pm}$ \eqref{M_pm_pm2}, the matrix of the dynamical system becomes degenerate
\[
A(M)=\displaystyle\left(\begin{array}{cccccc}
0 & 1 & 0 & 0 &0 & 0 \\
0 & 0 & 0 & 0 & 0 & 0\\
0 & 0 & 0 & 1 & 0 & 0 \\
0 & 0 & 2m^2 & 0 & 0 & 0\\
0 & 0 & 0 & 0 & 0 & 1 \\
0 & 0 & 0 & 0 & -2\mathfrak{m}^2 & 0 \\
\end{array}\right)_{M{\infty,\pm,\pm}},
\]
and has the following eigenvectors, respectively, pairs of coordinates $[\xi,H]$, $[\Phi,Z]$ and $[\varphi,z]$:
\begin{equation}\label{I-k_i}
k_{1,2}=0;\quad k_{3,4}=\pm \sqrt{2}m;\quad k_{5,6}=\pm i\sqrt{2}\mathfrak{m}.
\end{equation}
These eigenvalues correspond to the fact that for $\xi\to+\infty$ the phase trajectories in the plane $[\Phi,Z]$ are unstable near the singular point $\alpha\Phi^2=m^2$ (saddle), but stable in the plane $[\varphi,z]$ near the singular point $\beta\varphi^2=\mathfrak{m}^2$ (attracting center).

At the points $M_{\infty,\pm,0}$ \eqref{M_pm_02} the matrix is equal to
\[
A(M)=\displaystyle\left(\begin{array}{cccccc}
0 & 1 & 0 & 0 &0 & 0 \\
0 & 0 & 0 & 0 & 0 & 0\\
0 & 0 & 0 & 1 & 0 & 0 \\
0 & 0 & 2m^2 & 0 & 0 & 0\\
0 & 0 & 0 & 0 & 0 & 1 \\
0 & 0 & 0 & 0 & \mathfrak{m}^2 & 0 \\
\end{array}\right)_{M{\infty,\pm,0}},
\]
and has the following eigenvectors respectively to pairs of coordinates $[\xi,H]$, $[\Phi,Z]$ and $[\varphi,z]$:
\begin{equation}\label{II-k_i}
k_{1,2}=0;\quad k_{3,4}=\pm \sqrt{2}m;\quad k_{5,6}=\pm \mathfrak{m}.
\end{equation}
These eigenvalues correspond to the fact that for $\xi\to+\infty$ the phase trajectories in the plane $[\Phi,Z]$ are unstable near the singular point $\alpha\Phi^2=m^2$ (saddle), as in the plane $[\varphi,z]$ near the singular point $\varphi=0$  (attracting center).

In the case of singular points $M_{\infty,0,\pm}$ \eqref{M_0_pm2}, we similarly obtain the following eigenvalues:
\begin{equation}\label{III-k_i}
k_{1,2}=0;\quad k_{3,4}=\pm i m;\quad k_{5,6}=\pm i\sqrt{2} \mathfrak{m}.
\end{equation}
Thus, the singular points $M_{\infty,0,\pm}$ are stable (attracting) centers, both in the plane $[\Phi,Z]$ ($\Phi=0$) and in the plane $[\varphi,z]$ ($\varphi=\pm \mathfrak{m}$). For $\xi\to+\infty$, the phase trajectories asymptotically tend to special points $M_{\infty,0,\pm}$.

We note the following important circumstance. Since, as we found out in the section \ref{asimp_dyn8}, the system of equations \eqref{dxi/dt2} -- \eqref{Surf_Einst2} for $\xi\to+\infty$ asymptotically tends to the system of equations for the vacuum Higgs doublet \eqref{dH/dt_Vac} -- \eqref{SurfEinst_Vac}, then for $\xi\to+\infty$, the dynamic system will have singular points in phase hyperspace

\[\mathbb{R}_5=\{H,\Phi,Z,\varphi,z\}\subset \mathbb{R}_6\]
singular points that completely coincide with the singular points of the vacuum asymmetric scalar doublet \cite{YuKokh_TMF}. Therefore, for $\xi\to+\infty$, all the properties of a vacuum asymmetric scalar doublet are valid for this system.

\section{Numerical simulation}
Further, to shorten the letter, we will set a set of fundamental parameters of the \MII model using an ordered list
\[\mathbf{P_2}=[[\alpha,m,e],[\beta,\mathfrak{m},\epsilon],\pi_0,\Lambda]\]
and the initial conditions are an ordered list
\[\mathbf{I}=[\Phi_0,Z_0,\varphi_0,z_0,\rho],\]
where $\rho=\pm1$, and the value of $\rho=+1$ corresponds to the non-negative initial value of the Hubble parameter $H_0=H_+\geqslant0$, and the value of $\rho=-1$ corresponds to the negative initial value of the Hubble parameter $H_0=H_-<0$. At the same time, using the autonomy of the dynamical system, we assume $\xi(0)=0$ everywhere. Thus, the \MII model is determined by 10 fundamental parameters and 5 initial conditions, the \MI\ model is determined by 11 parameters $\mathbf{P_1}=[[\alpha,m,e,m_c,\pi_c],[\beta,\mu,\epsilon,m_f,\pi_f],\Lambda]$ and 5 initial conditions, the \MOO model is determined by 5 fundamental parameters $\mathbf{P}_{00}=$ $[\mu,\mathbf{e},$ $q,m,\pi_0,\Lambda]$ and one indicator $\mathbf{e}=\pm1$, where $\mathbf{e}=+1$ corresponds to a canonical field, $\mathbf{e}=-1$ is a phantom field. The initial conditions for this model are set by a list of two elements $\mathbf{I}_{00}=[\Phi_0,Z_0]$.
\subsection{Relation of the \MII\ model to previously studied models for scalar singlets}
In the case when $\xi(t)$ is a monotonically increasing function, i.e., $H\geqslant0$, the behavior of all models is practically the same. Let us consider the behavior of models for the case of a canonical scalar field under the following parameters and initial conditions:
\begin{eqnarray}\label{param0}
\mathbf{P_0}=[[1,1,1],[1,1,0],0.1,0.02];\\
\label{IC0}
\mathbf{I_0}=[0.2,0.1,0,0,1].
\end{eqnarray}

The evolution of the function $\xi(t)$ and the Hubble parameter $H(t)$ for this case is shown in Fig. \ref{xi_H_plus} and Fig. \ref{H_H_plus}. The graphs for the \MI\ and \MII\ models are the same -- the dashed lines on these graphs merge with the bold lines, which is to be expected (see the section \ref{sub_srav}). Some difference in the graph for the \MOO\ model is caused by the difference between the quadratic potential and the Higgs potential. We have specifically considered the different potentials in these models to demonstrate the influence of their shape on cosmological evolution. For $ \alpha=0$ both models give the same results in the area of $H\geqslant0$. Therefore, it can be argued that with the same parameters that ensure the non-negativity of the Hubble parameter $H\geqslant0$, all four models under the conditions of the section \ref{sub_srav} give the same results.

The output of the Hubble parameter to the horizontal line, shown in Fig. \ref{H_H_plus}, according to \eqref{Omega} means that in this case $\Omega(t\to+\infty)\to 1$ and $w=-1$, i.e., over time, models with such parameters go to inflation.
\Fig{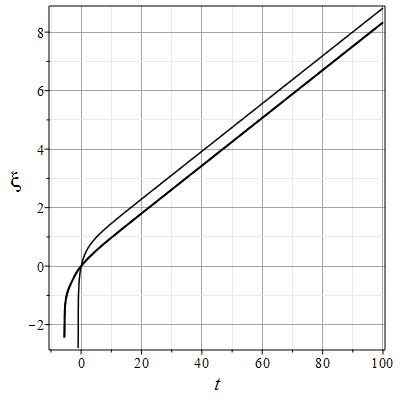}{7}{\label{xi_H_plus} Evolution of the function $\xi(t)=\ln(a(t))$ in various models with a single canonical field $\Phi$ under the parameters \eqref{param0} and the initial conditions \eqref{IC0}: the bold line is the \MII model, the dashed line is the \MI model, the thin line is the \MOO model.}
\Fig{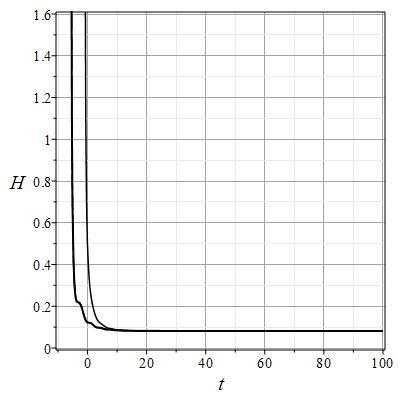}{7}{\label{H_H_plus} Evolution of the Hubble parameter $H(t)$ in various models with a single canonical field under the parameters \eqref{param0} and the initial conditions \eqref{IC0}: the bold line is the \MII model, the dashed line is the \MI model, the thin line is the \MOO model.}

However, in cases where the non-negativity condition of the Hubble parameter $H\geqslant0$ is violated, the behavior of the \MII\ and \MOO\ models is very different. The behavior of the \MII\ and \MOO\ models in the case of a phantom singlet is shown in Fig. \ref{Ris_Hf_0-2}. It can be seen from this graph that the evolution of the Hubble parameter $H(t)$ in the \MOO\ model is limited to the domain $H\geqslant0$ and clearly reveals the incorrectness of this model. In the \MII\ model the Hubble parameter starts with a constant asymptotic $H(-\infty)=-0.3$ and eventually reaches the symmetric asymptotic $H(+\infty)=+0.3$. These values $H=\pm0.3$ are just correspond to the attracting singular points in the model of a vacuum phantom singlet (see \cite{YuKokh_TMF}). Thus, in this case, the inflationary compression ($H<0,\Omega=1$) in the past is replaced by an inflationary expansion ($H>0,\Omega=1$) in the future.

This example, among other things, shows that although from the point of view of qualitative theory, the dynamical system \MII\ in this case, strictly speaking, has no singular points in the phase space $\mathbb{R}_6$, but, in fact, due to the asymptotic vacuum character at $\xi\to\infty$, the dynamical subsystem in $\mathbb{R}_5$ \emph{asymptotically} becomes autonomous and has singular points\footnote{We talked about that in the section \ref{asimp_dyn8}}.
\Fig{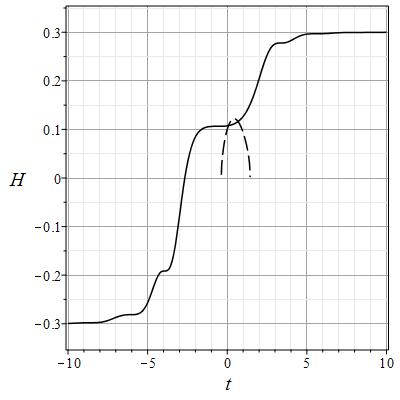}{8}{\label{Ris_Hf_0-2} Evolution of the Hubble parameter $H(t)$ in various models with a single canonical field under the parameters \eqref{param0} and the initial conditions \eqref{IC0}: the bold line is the \MII model, the dashed line is the \MOO model.}

\subsection{Oscillatory regimes of cosmological expansion: $\Lambda<0$}
In the case $\Lambda<0$ in the \MII model, as in the \MI model, oscillatory regimes of cosmological expansion are possible. Their implementation requires components of the scalar doublet that are approximately equal in order of magnitude.
\Fig{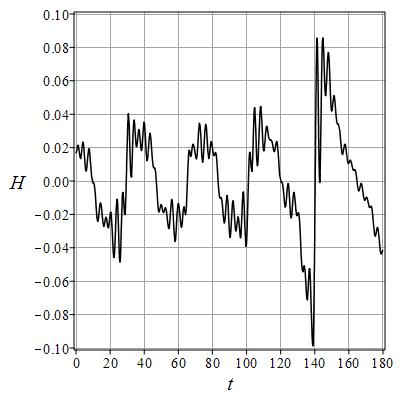}{7}{\label{Ris_Ht_2} Evolution of the Hubble parameter $H(t)$ of the \MII\ model for the parameters $\mathbb{R}_\Phi=\{\Phi,Z,H\}$: $\mathbf{P}=\mathbf{P}_{1}$; initial conditions $\mathbf{I}_1$.}

The Fig. \ref{Ris_Ht_2} shows a typical example of the oscillatory mode of the model \MII\ for the parameters of the model $\mathbf{P}_{1}=[[1,1,1],[1,0.5,1],0.01,-0.02]$ and the initial conditions $\mathbf{I}_1=[0.2,0.1,0.1,0.1,1]$. Fig. \ref{Ris_FZH_2} and \ref{Ris_varzH_2} show three-dimensional projections of the phase trajectories of the system.

It should be noted that the oscillatory behavior of the \MII\ model is very close to the behavior of the \MI\ \cite{Yukoch_TMF} model, as well as models with a vacuum asymmetric scalar doublet \cite{YuKokh_TMF}.
\Fig{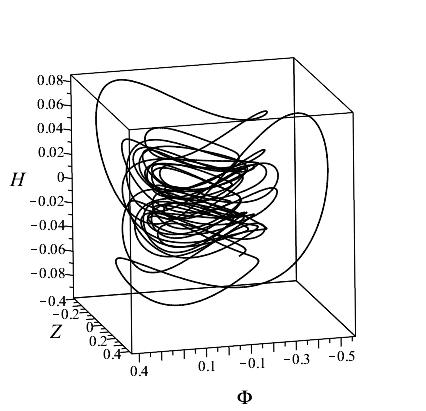}{8}{\label{Ris_FZH_2} Phase trajectory of the \MII\ model in the projection $\mathbb{R}_\Phi=\{\Phi,Z,H\}$: $\mathbf{P}=\mathbf{P}_{1}$; initial conditions $\mathbf{I}_1$.}
\Fig{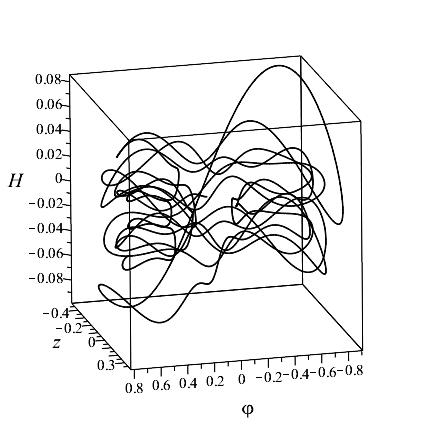}{8}{\label{Ris_varzH_2} Phase trajectory of the \MII\ in the projection $\mathbb{R}_\varphi=\{\varphi,z,H\}$: $\mathbf{P}=\mathbf{P}_{1}$; initial conditions $\mathbf{I}_1$.}

\subsection{An example of a cosmological model with a finite history}
As follows from the results of the section \ref{ssM1} in systems of scalar charged fermions, it is possible to achieve a cosmological singularity $a(t_1)\to0$ in the expansion phase of $H(t_1)\to+\infty$ and in the compression phase $a(t_2)\to0$, $H(t_2)\to-\infty$. Such a universe seems to exist for a limited time $t_2-t_1$. Note that in the incomplete model \MOO due to the non-negativity of the Hubble parameter, the universe exists for an infinite time.

Let's consider a numerical model of such a process with the parameters $\mathbf{P}_2= [[0,0,0.00001],[0,10^(-8),0.00001,0],0.1,0]$ and the initial conditions of $\mathbf{I}_2=[0,0,5*10^(-8),0,1]$. This case corresponds to a scalar phantom singlet.

The behavior of the scale factor with the same parameters is almost the same in the \MII\ and \MI\ models (see \cite{Ignat20_2}).
\section{Generating scalar doublet components in the \MII model}
In the examples above, we have demonstrated the coincidence of the behavior of the \MI\ and \MII\ models in the cases of scalar singlets. Therefore, it may seem that these models do not actually differ from each other. However, this is not the case. The fundamental difference is as follows. The scalar charge densities, $\sigma_c$ and $\sigma_f$, (formulas \eqref{sigma-sigma}) contain the same factor $(e\Phi+\epsilon\varphi)^3$, so even if one of the scalar fields is absent in the doublet, the scalar charge density for this component turns out to be different from zero, which leads to the generation of this field. We will demonstrate this phenomenon with examples.

\Fig{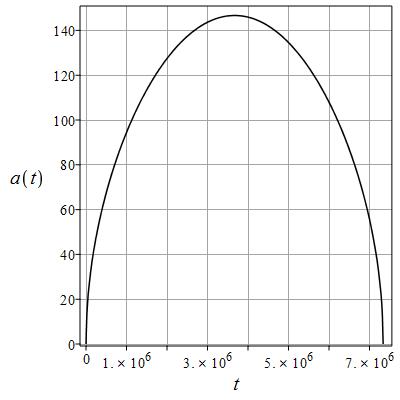}{7}{\label{Ris_At_21} Evolution of the scale factor $a(t)$ for the model \MII\ for the parameters $\mathbf{P}_2$.}
\subsection{Zero initial values for a phantom scalar field}
Let's set zero initial conditions for the phantom field $\varphi(0)=0;z(0)=0$:
\begin{eqnarray}\label{I3}
\mathbf{P}_3=[[1,1,1],[1,0.5,1],0.1,0.02];\\
\label{P3}
\mathbf{I}_3:=[0.01,0,0,0,1].
\end{eqnarray}
The Fig. \ref{Ris_varF_gen1t_2} and \ref{Ris_F_gen1t_2} show the evolution of scalar potentials under zero initial conditions for a phantom scalar field.

In this case, the canonical scalar field, in fact, is present only on a small time interval of the order of $\Delta \sim 70$ Planck times, performing several oscillations on this interval. Therefore, in this case we are dealing with the generation of the canonical scalar field of the phantom component of the doublet. In this case, the dynamic system starts from a stable singular point of the vacuum asymmetric scalar doublet with a transition to another stable singular point of the vacuum asymmetric scalar doublet.
As we noted above, the choice of the ``initial'' time value $t=0$ is relative, and the graphs in Fig. \ref{Ris_varF_gen1t_2} and \ref{Ris_F_gen1t_2} convincingly demonstrate this fact. It turns out that the potential of the phantom field $\varphi$ was constant in the past, as well as in the future $\varphi(\mp\infty)=\mp0.5$ and, again, coincides with the stable singular points of the asymmetric vacuum scalar doublet \cite{YuKokh_TMF}. It turns out that the potential of the canonical field $\Phi$ both in the past and in the future tends to zero $\Phi(\mp\infty)=0$, i.e., it starts from a stable zero singular point and then returns to it\footnote{The results of numerical integration give $\Phi(\mp\infty)=10^{-15}\div10^{-14}$.}.
\Fig{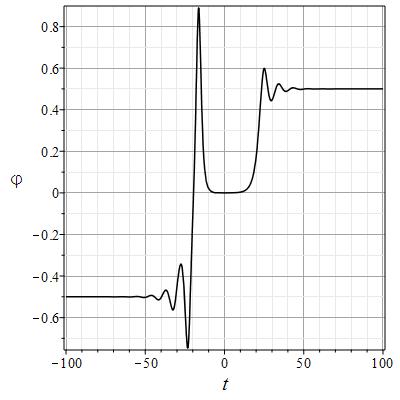}{7}{\label{Ris_varF_gen1t_2} Evolution of the phantom potential $\varphi(t)$ for the \MII\ model for the parameters \eqref{P3} and the initial conditions \eqref{I3}.}
\Fig{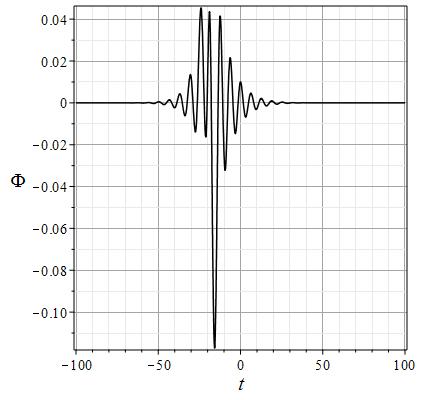}{7}{\label{Ris_F_gen1t_2} Evolution of the canonical potential $\Phi(t)$ for the \MII\ model for the parameters \eqref{P3} and the initial conditions \eqref{I3}.}

The Fig. \ref{Ris_xi_gen1t_2} and \ref{Ris_Hi_gen1t_2} show the graphs of the evolution of the scale function $\xi(t)$ and the Hubble parameter $H(t)$ for the case under study. These graphs clearly show that asymptotically vacuum stable states correspond to the phases of cosmological compression and expansion, a fairly rapid transition between which is accompanied by bursts of the canonical scalar field. We can say that a dynamic system with the help of a canonical field sends some characteristic signal about the transition between the stages of compression and expansion.
\Fig{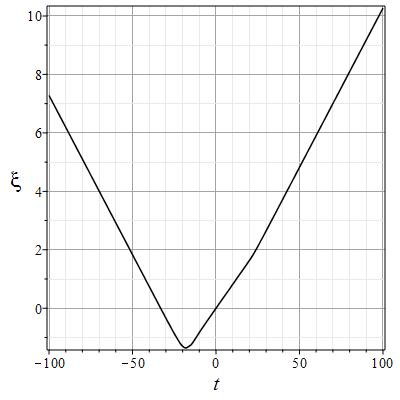}{8}{\label{Ris_xi_gen1t_2} Evolution of the scale function $\xi(t)=\ln a(t)$ for the model \MII\ for the parameters \eqref{P3} and the initial conditions \eqref{I3}.}
\Fig{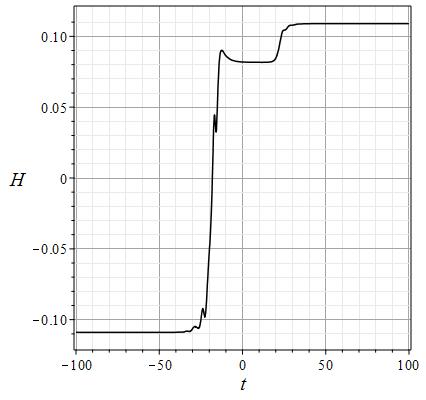}{8}{\label{Ris_Hi_gen1t_2} Evolution of the Hubble parameter $H(t)$ for the \MII\ model for the parameters \eqref{P3} and the initial conditions \eqref{I3}.}
\subsection{Zero initial values for a canonical scalar field}
Let's set zero initial conditions for the canonical field $\Phi(0)=0;Z(0)=0$, while preserving the values of the parameters $\mathbf{P}_3$:
\begin{eqnarray}\label{I4}
\mathbf{I}_4:=[0,0,0.00001,0,1].
\end{eqnarray}
The Fig. \ref{Ris_F_gen2t_2} and \ref{Ris_varF_gen2t_2} show the evolution of scalar potentials under zero initial conditions for a canonical scalar field.
\Fig{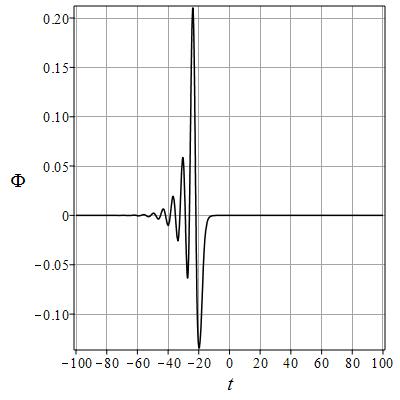}{8}{\label{Ris_F_gen2t_2} Evolution of the canonical potential $\Phi(t)$ for the \MII\ model for the parameters \eqref{P3} and the initial conditions \eqref{I4}.}
\Fig{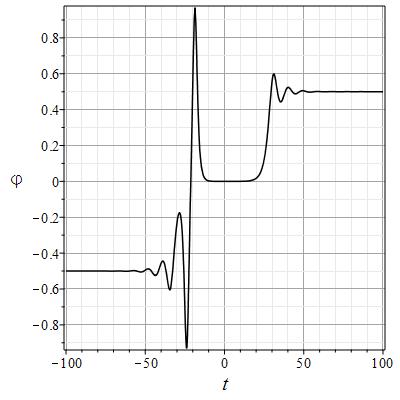}{8}{\label{Ris_varF_gen2t_2} Evolution of the phantom potential $\varphi(t)$ for the \MII\ model for the parameters \eqref{P3} and the initial conditions \eqref{I4}.}
We see that the situation does not differ qualitatively from the one discussed in the previous section: a dynamic system starts from a stable point of an asymptotically vacuum state of a scalar doublet with a zero canonical scalar field (compression phase) and passes to another symmetric stable point (expansion phase). Thus, we can conclude that it is the canonical scalar field that is generated by the phantom field through the effective charge of the Fermi system.
\subsection{Energy picture of the process of generating a canonical scalar field}
It was noted in \cite{TMF_print21} that systems of scalar charged particles with zero seed mass have a unique property: they allow a scalar neutral state in which scalar fields are completely absent in the presence of arbitrary scalar charges of particles. In this case, the density of the scalar charge also vanishes, and the particles become massless. Thus, the ultrarelativistic state of a plasma with zero scalar fields becomes the main state of such a statistical system. This property of systems of scalar charged particles radically distinguishes them from systems with electromagnetic interaction.

Due to this property and the above-mentioned proportionality of the density of scalar charges $\sigma_c$ and $\sigma_f$ to the factor $(e\Phi+\epsilon\varphi)$ each of the components of an asymmetric scalar doublet plays the role of a catalyst for generating the second component. As we saw above, it is the phantom field that is the catalyst for generating the classical one. However, as the analysis shows, the influence of the phantom field on the classical generation process is not limited to the role of the catalyst, but is significantly determined by the energy of this field and the energy of fermions during the transition from the compression phase to the expansion phase.

The Fig. \ref{Ris_Ec_Ef} and \ref{Ris_Es_Ep} show graphs of the evolution of the energy densities of individual components of a dynamical system for the parameters \eqref{P3} and the initial conditions \eqref{I4} considered in the previous section. It can be seen from these graphs that the process of generating a canonical scalar field is characterized, firstly, by a simultaneous burst of the negative energy density of the phantom field and the positive energy density of fermions, and, secondly, by the zero effective total energy $\varepsilon_{e\!f\!f}$.

\section*{Conclusion}
Thus, first, we conducted a qualitative analysis of the cosmological model based on a one-component system of doubly scalar charged degenerate fermions, and investigated the asymptotic and limiting properties of this model. Secondly, we built an appropriate numerical model with the help of which we conducted a comparative analysis of this model with previously studied models. Third, we have identified the possibility of generating a canonical scalar field in this model by a phantom scalar field by means of a canonical scalar charge density self-induced by this field.

Summing up the results of this study, we will list its most important.\\
\noindent$\bullet$ A closed mathematical model of a cosmological system consisting of a one-component system of degenerate doubly scalar charged fermions and an asymmetric pair of scalar fields, canonical $\Phi$ and phantom $\varphi$, with Higgs potential energy, is formulated. It describes a dynamical system in a 6-dimensional phase space $\mathbb{R}_6=\{\xi,H$, $\Phi,Z=\dot{\Phi}$, $\varphi,z=\dot{\varphi}\}$. The dynamical system is completely described by a normal autonomous system of ordinary differential equations with respect to the cosmological time $t$. Its behavior in a model with a cosmological term is determined by 11 parameters.\\

\Fig{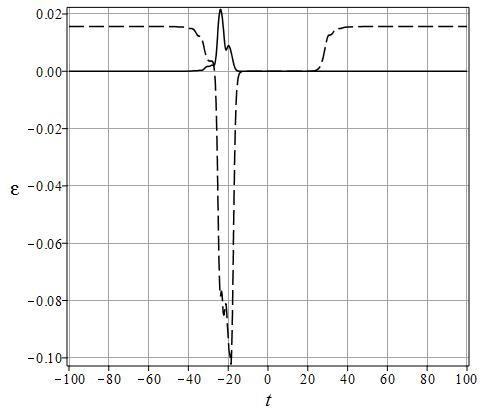}{8}{\label{Ris_Ec_Ef} Evolution of the energy densities of the classical $\varepsilon_c(t)$ (solid line) and the phantom $\varepsilon_f$ (dashed line) scalar field for the parameters \eqref{P3} and the initial conditions \eqref{I4}.}
\Fig{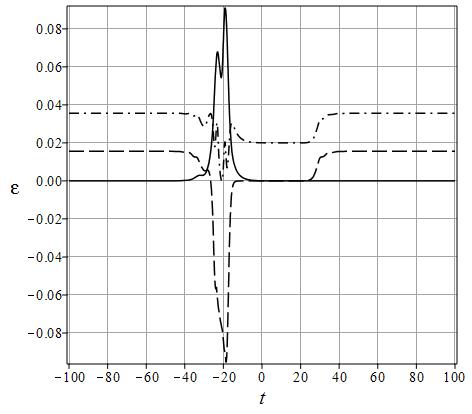}{8}{\label{Ris_Es_Ep} Evolution of the total energy density of the scalar field $\varepsilon_s (t)$ (dashed line), the energy density of fermions $\varepsilon_p(t)$ (solid line) and the total effective energy density of $\varepsilon_s\!\!+\!\varepsilon_p\!\!+\!\!\Lambda$ (dashed dashed line) for the parameters \eqref{P3} and the initial conditions \eqref{I4}.}

\noindent$\bullet$ The connection of this model with the previously studied models based on a one-component system of degenerate singly scalar charged degenerate fermions and a two-component system of variously scalar charged degenerate fermions is established.\\
\noindent$\bullet$ A qualitative analysis of the dynamical system corresponding to the presented cosmological model is carried out, the character of the singular points corresponding to an infinite future asymptotic vacuum asymmetric scalar doublet is established.\\
\noindent$\bullet$ The asymptotic and limiting properties of the models are investigated. It is shown that in cases admitting an infinite expansion of the universe, its asymptotic properties at infinity are completely determined by vacuum scalar fields, i.e., mainly by late inflation. At the same time, the model also admits finite histories of the universe, in these cases, the exit from the singular state and the entry into it occurs according to the asymptotics of the ultrarelativistic model. Asymptotically exact solutions of the dynamical system corresponding to the entry into and exit from the singular state are found.\\
\noindent$\bullet$ The behavior of the studied cosmological model is compared by numerical integration methods with the previously studied incomplete models of the cosmological evolution of charged degenerate fermions. The coincidence of the model properties with the previously studied models in the limiting cases of parameters is shown.\\
\noindent$\bullet$  A unique property of the presented model is revealed -- the possibility of generating a canonical scalar field by a phantom scalar field by self-induction of a scalar charge density. The mechanism of generation of a standard scalar field by self-induction of a scalar charge near the moment of phase change of cosmological compression and cosmological expansion is investigated.

Taking into account the results of the study of the gravitational stability of scalar charged plasma \cite{GC_21_2}, as well as the results of \cite{Ignat_Sasha_G&G} regarding the possible increase in the effective mass of scalar charged fermions in the presence of a phantom field, the analysis carried out in this article shows the prospects of the proposed model of a one-component system of doubly scalar charged fermions as a possible model for explaining the mechanism of formation of supermassive objects in the early Universe. To implement this idea, it is necessary, first of all, to study this model for gravitational stability, which we intend to do in the near future.

\subsection*{Funding}

This work was funded by the subsidy allocated to Kazan Federal University for the state assignment in the sphere of scientific activities.
 

\end{document}